\documentclass[12pt]{article}
\def\ypiz_2gg{$\pi^0 \rightarrow \gamma\gamma$}

\begin {document}
\title
{Search for $\Theta^+(1540)$ emission in hadron--nucleus
collisions at 400--700 GeV}
\author{ A. E. Asratyan and V. A. Matveev \\
{\normalsize \it Institute of Theoretical and Experimental Physics, Moscow, 117218 Russia} \\
{\normalsize Email: ashot.asratyan@gmail.com, matveev@itep.ru} }
\date {\today}
\maketitle

\begin{abstract}
The data on hadron--nucleus collisions at 400--700 GeV, collected by the SELEX 
experiment at Fermilab, are analyzed for formation of the exotic pentaquark 
baryon $\Theta^+(1540)$. A narrow enhancement near 1539 MeV is observed in the 
mass spectrum of the $pK^0_S$ system emitted at small $x_F$ from hadron 
collisions with copper nuclei. The statistical significance of the peak is 
near 9 standard deviations. Fitted width of the observed $pK^0_S$ resonance is 
consistent with being entirely due to experimental resolution, and its 
intrinsic width is restricted to $\Gamma < 3$ MeV at 90\% CL. The data favor 
positive rather than negative strangeness for the $pK^0_S$ resonance observed 
in $h$Cu collisions. At the same time, the $pK^0_S$ mass spectrum for 
collisions in carbon is featureless. The yield of $\Theta^+$ baryons per $h$C 
collision is restricted to be $< 24$\% of the yield per $h$Cu collision.
\end{abstract}

\newpage

     Possible existence of multiquark hadrons, and pentaquark baryons
in particular, has been discussed ever since the quark model was 
proposed \cite{forerunners}. Fairly definite predictions 
for the antidecuplet of pentaquark baryons with spin--parity ${1/2}^+$ 
have been formulated by Diakonov, Petrov, and Polyakov in the framework 
of the chiral quark--soliton model \cite{DPP}. Their crucial prediction
has been that the explicitly exotic baryon with $S = +1$ and $I = 0$, 
the $\Theta^+(uudd\bar{s})$ that should decay to $nK^+$ and $pK^0$,
is relatively light and narrow: $m \approx 1530$ MeV and 
$\Gamma < 15$ MeV. More recent theoretical analyses suggest that the 
$\Theta^+$ intrinsic width may be as small as $\sim 1$ MeV or even 
less \cite{width}. Narrow peaks near 1540 MeV in the $nK^+$ and $pK^0$ 
mass spectra were initially detected in low-energy photoproduction by 
LEPS \cite{Nakano-2003} and in the charge-exchange reaction 
$K^+n \rightarrow pK^0$ by DIANA \cite{DIANA-2003}. Yet another early 
observation in the $pK^0$ channel, that endured the time test, relied on 
combined bubble-chamber data on $\nu N$ collisions \cite{Neutrino}.
Subsequently,
both LEPS and DIANA were able to confirm their initial observations 
\cite{Nakano-2009, DIANA-2007, DIANA-2010, DIANA-2014, DIANA-2015}. 
Using the unique properties of the charge-exchange reaction that forms 
$\Theta^+$ baryons in the $s$-channel, DIANA reported a direct measurement 
of the $\Theta^+$ intrinsic width: $\Gamma = 0.34\pm0.10$ MeV assuming 
$J = 1/2$ \cite{DIANA-2014}. 

     Other searches for the $\Theta^+$ baryon in 
different reactions and experimental conditions yielded positive evidence as 
well as null results casting doubt on its existence, see the review papers 
\cite{Burkert, Danilov-Mizuk, Hicks}. One of the heftiest null results
was reported by CLAS using a large sample of $\gamma p$ collisions \cite{CLAS}.
However, the same sample was later reanalyzed in terms of $\phi p$ and
$\Theta^+ \bar{K^0}$ interference in the $p K^0_S K^0_L$ final state, and a
statistically significant peak near 1539 MeV was found in the $K^0_S$ missing 
mass \cite{Amaryan}. Of the multitude of null results, only a few that are 
formulated in terms of the $\Theta^+$ intrinsic width $\Gamma$ should be treated
as physically meaningful. The best upper limit to date, $\Gamma < 0.36$ MeV 
assuming $J^P = 1/2^+$, was obtained in \cite{E19} by analyzing the $K^-$
missing mass in the $\pi^- p \rightarrow K^- X$ reaction. This restriction is 
narrowly consistent with the DIANA measurement. The bulk of null results can 
be probably explained by the extreme smallness of the $\Theta^+$ width 
that implies the smallness of production cross-sections \cite{Diakonov}. 

     The properties of the pentaquark baryon $\Theta^+(1540)$ can be 
best probed in low-energy exclusive processes such as the
charge-exchange reaction $K^+n \rightarrow pK^0$. In high-energy 
collisions, Azimov {\it et al.} argue that the $\Theta^+$ 
baryon should be primarily formed from multiquark configurations of the 
target residue \cite{Azimov-Goeke-Strakovsky}, and therefore emitted with 
small $x_F$. In this report, we investigate the $pK^0_S$ mass
spectrum in SELEX, paying special attention to the kinematic region of
target fragmentation.

     The SELEX experiment operated in the Fermilab hyperon beam, and
was primarily designed for studying hadroproduction of charmed 
baryons and mesons in the forward region. The negative beam at 
$\langle p_\mathrm{beam}\rangle = 615$ GeV 
was composed of about 50\% $\Sigma^-$
and 50\% $\pi^-$. The positive beam at 
$\langle p_\mathrm{beam}\rangle = 540$ GeV
was 92\% protons. The beam spread was $\Delta p / p \simeq 8$\% HWHM.
A beam Transition Radiation Detector identified each beam particle as
meson or baryon with zero overlap. The beam interacted in a set of
five target foils ( 2 Cu and 3 C) spaced by 1.5 cm. Target thickness
was 1.06\% and 0.76\% $\lambda_\mathrm{int}$ for the two copper targets,
and 0.82\% $\lambda_\mathrm{int}$ for each of the three carbon targets.
As part of the triggering system, one scintillation counter was placed
upstream and two---downstream of the target assembly.
The three-stage magnetic spectrometer is shown
elsewhere \cite{spectrometer}. The most important features are the
high-precision, highly redundant vertex detector that provides an
average proper-time resolution of 20 fs for the charm decays, a
10-m-long Ring-Imaging Cherenkov (RICH) datector that separates 
$\pi$ from $K$ up to 165 GeV/c \cite{RICH}, and a high-resolution 
tracking system that has momentum resolution of $\sigma_p / p < 1$\% 
for a 150-GeV proton.
The triggering scheme was optimized for selecting charm candidate
events. A scintillator trigger demanded an inelastic collision with
at least four charged tracks in the interaction scintillators and at 
least two hits in the positive-particle hodoscope after the second 
analyzing magnet. Event selection in the online filter required full
track reconstruction for measured fast tracks ($p > 15$ GeV). These
tracks were extrapolated back into the vertex silicon planes and
linked to silicon hits. The beam track was measured in upstream
silicon detectors. A full three-dimensional vertex fit was then
performed. An event was logged on tape if any of the fast tracks in
the event were {\it inconsistent} with having come from a single primary
vertex. This filter passed 1/8 of all interaction triggers. The 
experiment recorded data from $15.2\times10^9$ inelastic interactions
and logged $1\times10^9$ events on tape using both positive and 
negative beams. The sample was 67\% $\Sigma^-$-induced, 14\%
$\pi^-$-induced, and 18\% proton-induced.

\begin{figure}[!t]

\vspace{ 10.5 cm}
\includegraphics{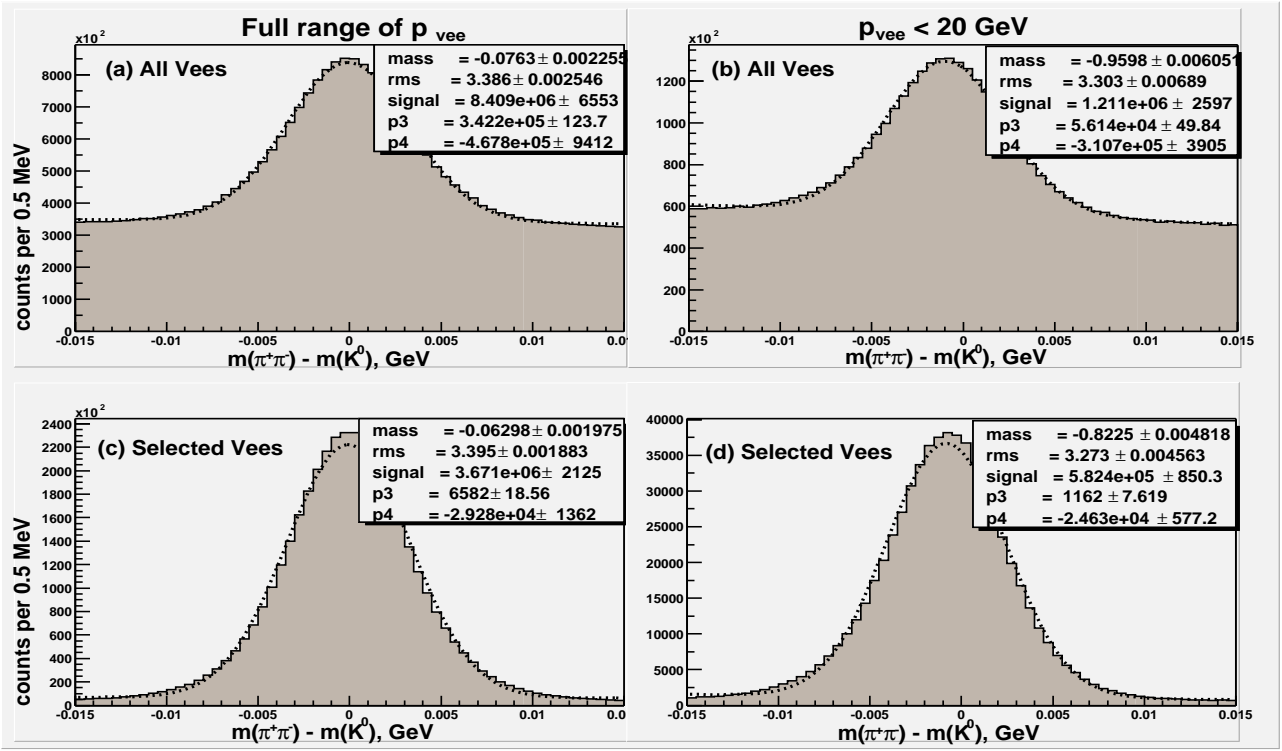}
\caption
{The effective $\pi^+\pi^-$ mass for all detected Vees prior to
analysis selections (a) and for the selected 
$K^0_S \rightarrow \pi^+\pi^-$  candidates (c).
The effect of the additional selection $p_\mathrm{vee} < 20$ GeV 
is illustrated in the right-hand panels (b) and (d).}
\label{kzero}
\end{figure}

     We analyze the complete SELEX dataset reconstructed using
standard routines plus some extra algorithms that allow to detect
more secondary vertices \cite{VBK}. In particular, additional Vee
particles that decayed inside the vertex detector are reconstructed by 
extrapolating spectrometer tracks back to the vertex detector
and matching them with hits in Silicon planes. Prior to analysis 
selections, we have some 8.4 million reconstructed 
$K^0_S \rightarrow \pi^+\pi^-$  
decays above large combinatorial background, see Fig.~\ref{kzero}a.
The reconstruction of pre-selected $\pi^+\pi^-$ decays of $K^0_S$
mesons with $p < 20$ GeV is separately illustrated in Fig.~\ref{kzero}b.
As soon as we are primarily interested in $pK^0_S$ pairs emitted in 
the kinematic region of target fragmentation, special attention must 
be paid to selecting relatively soft $K^0_S \rightarrow \pi^+\pi^-$  
and proton candidates. The selections used in this analysis are
described below.

     The primary vertex of the collision is found using the standard
algorithm. The beam track should be accurately reconstructed 
($\chi^2 < 9$), and is always retained when iteratively fitting the
primary vertex. For the latter, we require $\chi^2 < 6$. The
fitted primary vertex should lie either inside target material or
within 500 $\mu$m of the target. (The uncertainty on the $z$-coordinate
of the primary vertex is $\sim 200$ $\mu$m on average.)
The number of secondary tracks fitted to the primary vertex is restricted 
to the interval $5 < N < 20$. This is to select deep-inelastic collisions
and, on the other hand, to suppress secondary interactions and reduce the 
occupancy of Silicon planes.
     
     The  $K^0_S \rightarrow \pi^+\pi^-$  decays are selected as follows :
\begin{itemize}
\item
$|\cos\Theta^*_\pi| < 0.75$ for the decay angle in the $K^0_S$ frame,
whereby any reflection from  $\Lambda \rightarrow p\pi^-$  is completely
removed and the softest decay pions in lab are rejected ;
\item
$L/\sigma_L > 6$ for the $K^0_S$ path before decay in error units, aimed
at rejecting tracks emitted from the primary vertex ;
\item
DCA $> 20$ $\mu$m  for the Distance of Closest Approach between the track
of either daughter pion and the beam track, further rejecting tracks from 
the primary vertex ;
\item
Either $\chi^2_\mathrm{pvtx} < 9$ or $\beta_\mathrm{miss} < 0.5$ mrad
for associating the Vee with the reconstructed primary vertex. Here,
$\beta_\mathrm{miss}$ is the ``acoplanarity" angle between the $K^0_S$  
decay plane and its path in lab (straight line between the primary and 
secondary vertices);
\item
$p(\pi^\pm) > 3$ GeV for lab momenta of both decay pions, as suggested
by the momentum threshold of the spectrometer ;
\item 
$\chi^2 < 9$ for the tracks of both daughter pions, and  $\chi^2 < 4$
for the reconstructed secondary vertex.
\end{itemize}
The combined effect of these selections on the 
$K^0_S \rightarrow \pi^+\pi^-$  signal is shown in Figs.~\ref{kzero}c
and \ref{kzero}d for the full $K^0_S$ momentum interval and for 
$p < 20$ GeV, respectively. In what follows, the 
$K^0_S \rightarrow \pi^+\pi^-$  decays are selected in the mass 
interval $|m(\pi^+\pi^-)-m(K^0)| < 8$ MeV, which corresponds to 
some 2.5$\sigma$. 

     Proton candidates are selected as : (a) tracks that produced 
Cherenkov rings, for which the proton probability is the highest
according to RICH data (PID = 5); (b) tracks with momenta below the
Cherenkov threshold for protons, that impinged on the RICH but have not 
been identified as either kaons, pions, muons, or electrons (PID = 0). 
The additional selections are as follows :
\begin{itemize}
\item
The Distance of Closest Approach between the proton and beam tracks 
is restricted to DCA $< 15$ $\mu$m ; 
\item
For the sub-threshold proton candidates, the RICH pion and muon 
probabilities must not exceed 70\%, and the number of hits in the
Transition Radiation Detector must not exceed 3 ;
\item
The proton track should be accurately reconstructed ( $\chi^2 < 6$)
so as not to degrade the resolution on the $pK^0_S$ effective mass.
\end{itemize}

     For the instrumental smearing of $m(\pi^+\pi^-)$ in the decay
$K^0_S \rightarrow \pi^+\pi^-$ to cancel out, the mass of the $pK^0_S$ 
system is estimated as 
$m'(pK^0_S) = m(p\pi^+\pi^-) - m(\pi^+\pi^-) + m(K^0)$. 
The Feynman variable is defined as $x_F = p^*_L / p^*_\mathrm{max}$, 
where  $p^*_L$  is $pK^0_S$ longitudinal momentum in the collision 
frame assuming a target nucleon at rest in lab, and  $p^*_\mathrm{max}$
is its maximum value that corresponds to zero-mass recoil against the 
$pK^0_S$ system. 

\begin{figure}[t]

\vspace { 6 cm }
\includegraphics{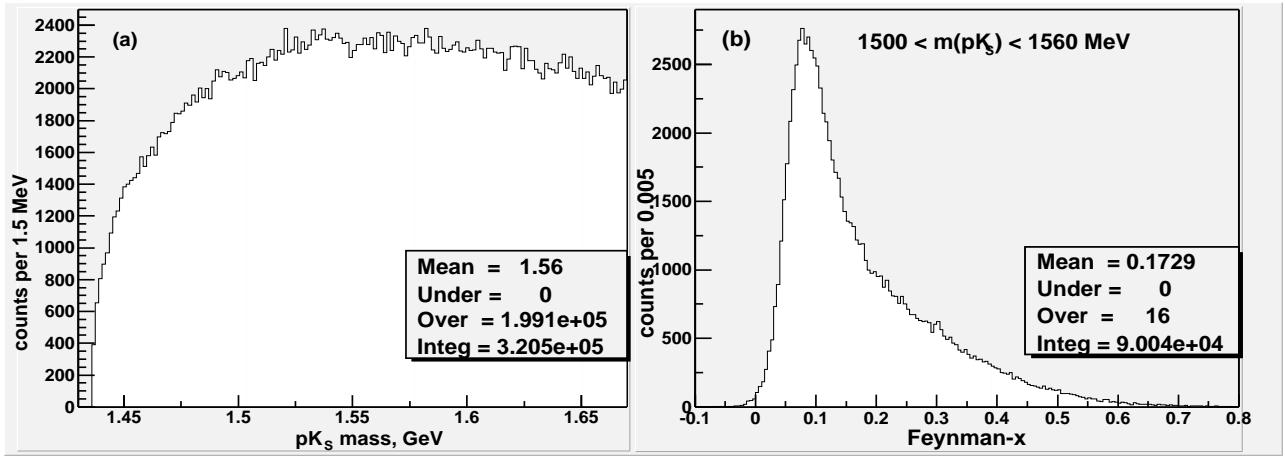}
\caption
{The $pK^0_S$ effective mass for all combinations of selected protons 
and $K^0_S$ mesons in all reconstructed events (a). Shown in (b) is 
the Feynman variable $x_F$ for selected $pK^0_S$ pairs with effective 
mass in the region of 1500--1560 MeV.}
\label{dimass-and-xF}
\end{figure}

\begin{figure}[t]

\vspace { 6 cm }
\includegraphics{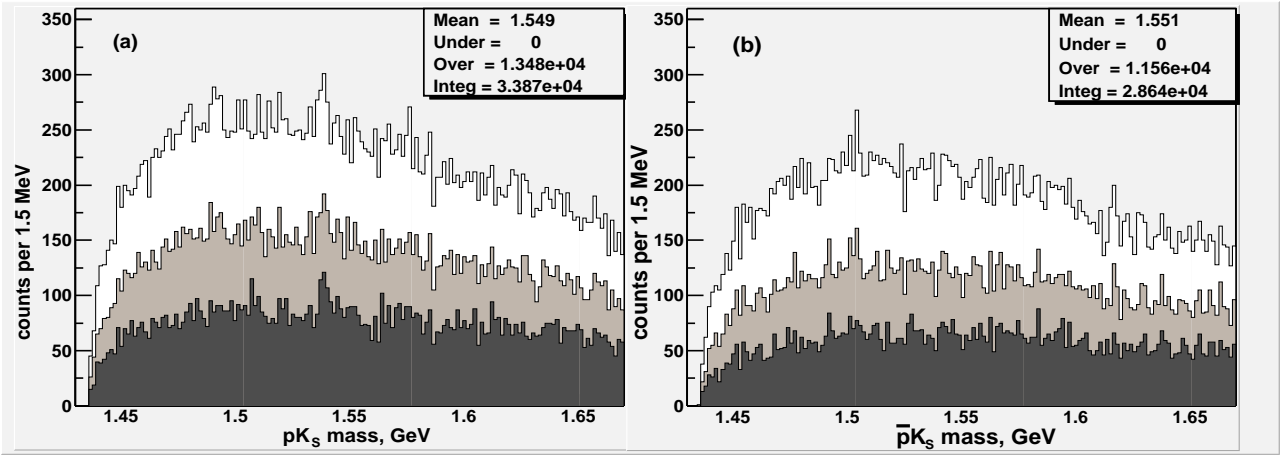}
\caption
{The $pK^0_S$ (a) and $\bar{p}K^0_S$ (b) mass spectra under the
selections $x_F < 0.06$, 0.05, and 0.04 (the open, shaded, and
dark histograms, respectively).}
\label{dimass-for-all-vary-xF}
\end{figure}

\begin{figure}[t]

\vspace { 5.7 cm }
\includegraphics{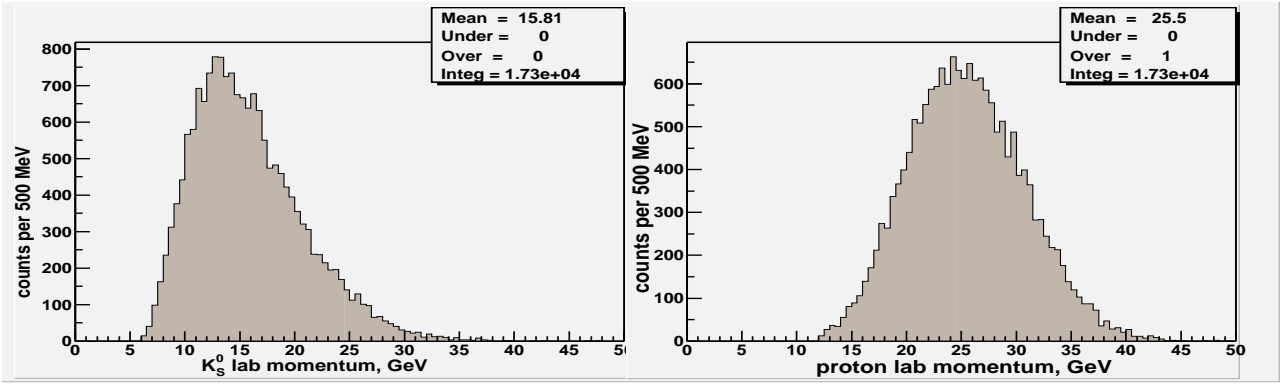}
\caption
{Lab momenta of those protons and $K^0_S$ mesons that form $pK^0_S$
pairs with $x_F < 0.04$.}
\label{momenta}
\end{figure}

     The $pK^0_S$ effective mass is plotted in 
Fig.~\ref{dimass-and-xF}a for all combinations of selected protons and
$K^0_S$ mesons in all reconstructed events. Here and in what follows, 
the data for the $\Sigma^-$, $\pi^-$, and proton projectiles have been 
combined. The $pK^0_S$ mass spectrum for the full range of $x_F$ shows 
no significant peaks in the mass region 1520--1540 MeV. The analogous 
mass spectrum for the $\bar{p}K^0_S$ system (not shown) is equally 
regular. That the $x_F$ distribution shown in Fig.~\ref{dimass-and-xF}b
steeply decreases towards $x_F = 0$ is the effect of detection
efficiency, {\it i.e.} of the losses of soft $K^0_S$ mesons and protons.
The $x_F$ distribution for the relevant mass region 
$1500 < m(pK^0_S) < 1560$ MeV is very similar to that obtained in 
the WA89 analysis of the $pK^0_S$ mass spectrum \cite{WA89-on-Theta}.
Note however that the kinematic region $x_F < 0.05$, that we find to 
be the most interesting, was not investigated in \cite{WA89-on-Theta}
for some unknown reason. The effect of cutting on $x_F$ is illustrated 
in Fig.~\ref{dimass-for-all-vary-xF} for either the $pK^0_S$ and
$\bar{p}K^0_S$ mass spectra. At $x_F$ values below some 0.05, 
an enhancement near 1540 MeV emerges in the $pK^0_S$ mass spectrum 
for collisions reconstructed in all targets, whereas the $\bar{p}K^0_S$ 
mass spectrum remains featureless. For the $pK^0_S$ pairs with 
$x_F < 0.04$, proton and $K^0_S$ lab momenta are plotted in 
Fig.~\ref{momenta}. Virtually all contributing protons are in the 
sub-threshold category (PID = 0) and, moreover, uncomfortably below 
the charged-kaon Cherenkov threshold of some 41 GeV.
% For the relatively soft $K^0_S$ mesons that pass the selection 
% $x_F < 0.04$, the uncertainty on the $z$-coordinate of the Vee
% vertex is $\sim 700$ $\mu$m on average.

\begin{figure}[t]

\vspace { 13.5 cm}
\includegraphics{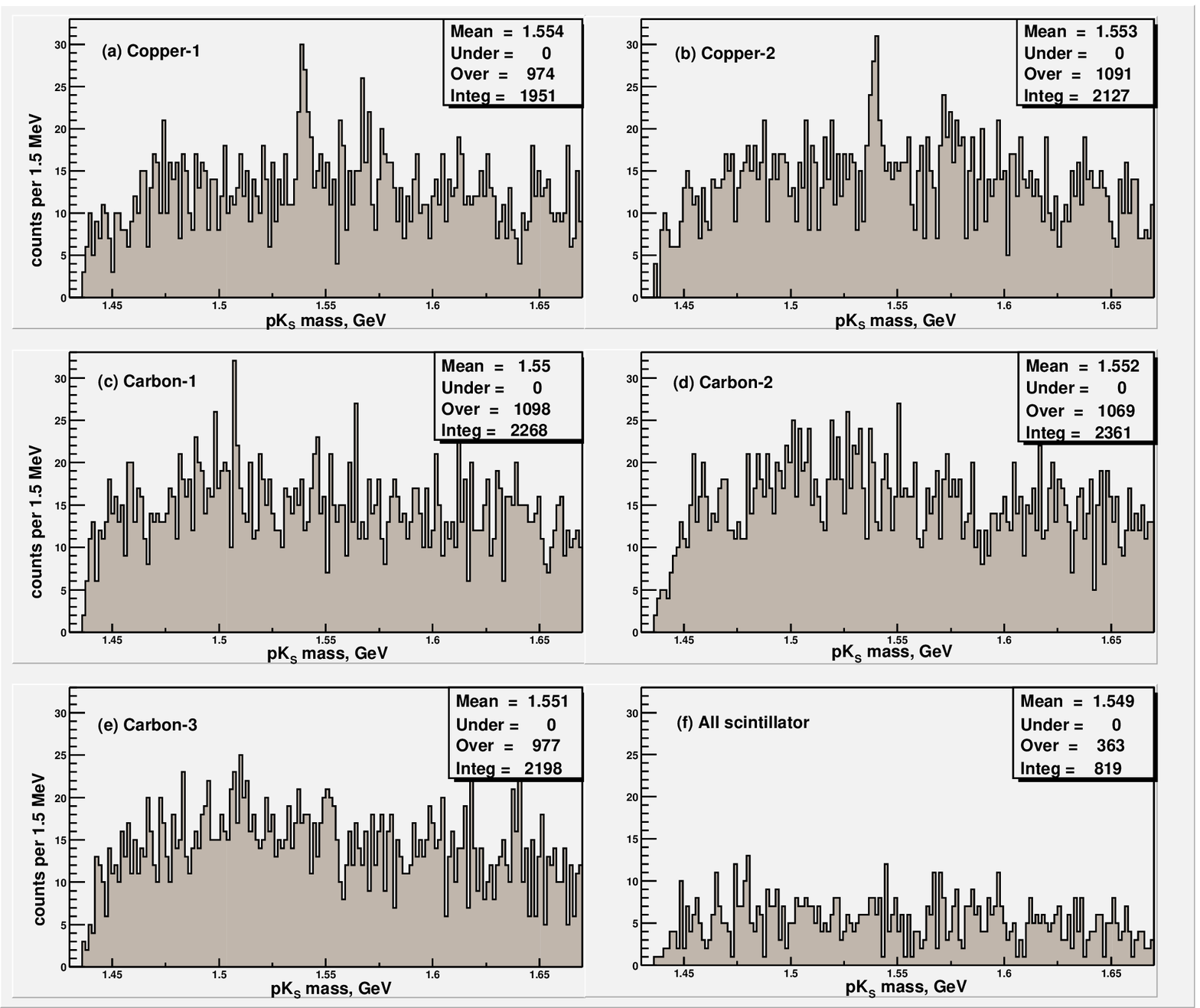}
\caption
{Breakdown of the $pK^0_S$ mass spectrum for $x_F < 0.04$ to five 
targets proper plus scintillator. The data for the three layers of
scintillator have been combined.}
\label{dimass-six-targets}
\end{figure}
\begin{figure}[t]

\vspace { 12 cm }
\includegraphics{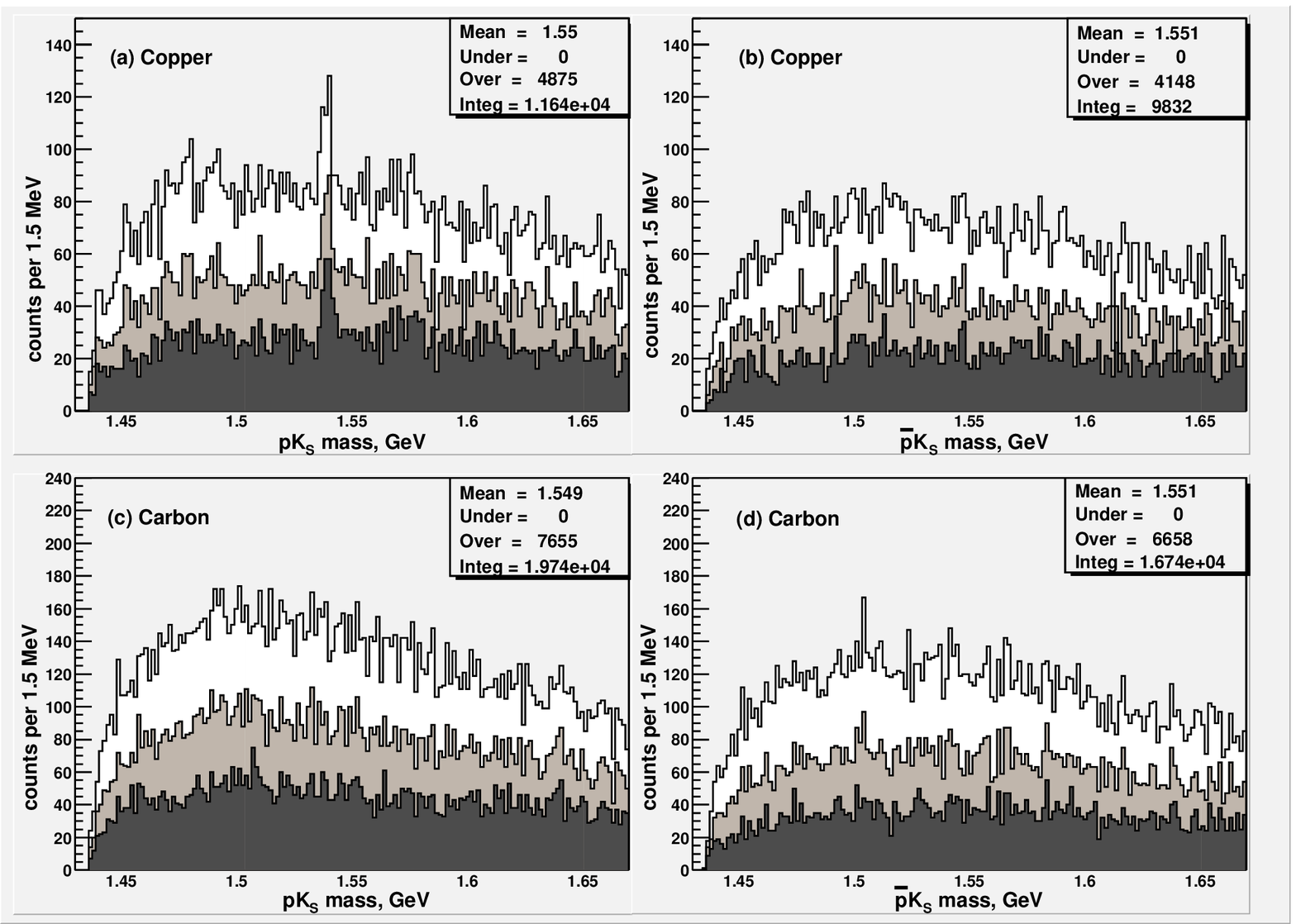}
\caption
{The $pK^0_S$ effective mass separately plotted for
collisions with copper and carbon under the selections
$x_F < 0.06$, 0.05, and 0.04 (the open, shaded, and 
dark histograms, respectively).}
\label{dimass-vary-xF}
\end{figure}
\begin{figure}[!t]

\vspace { 16 cm }
\includegraphics{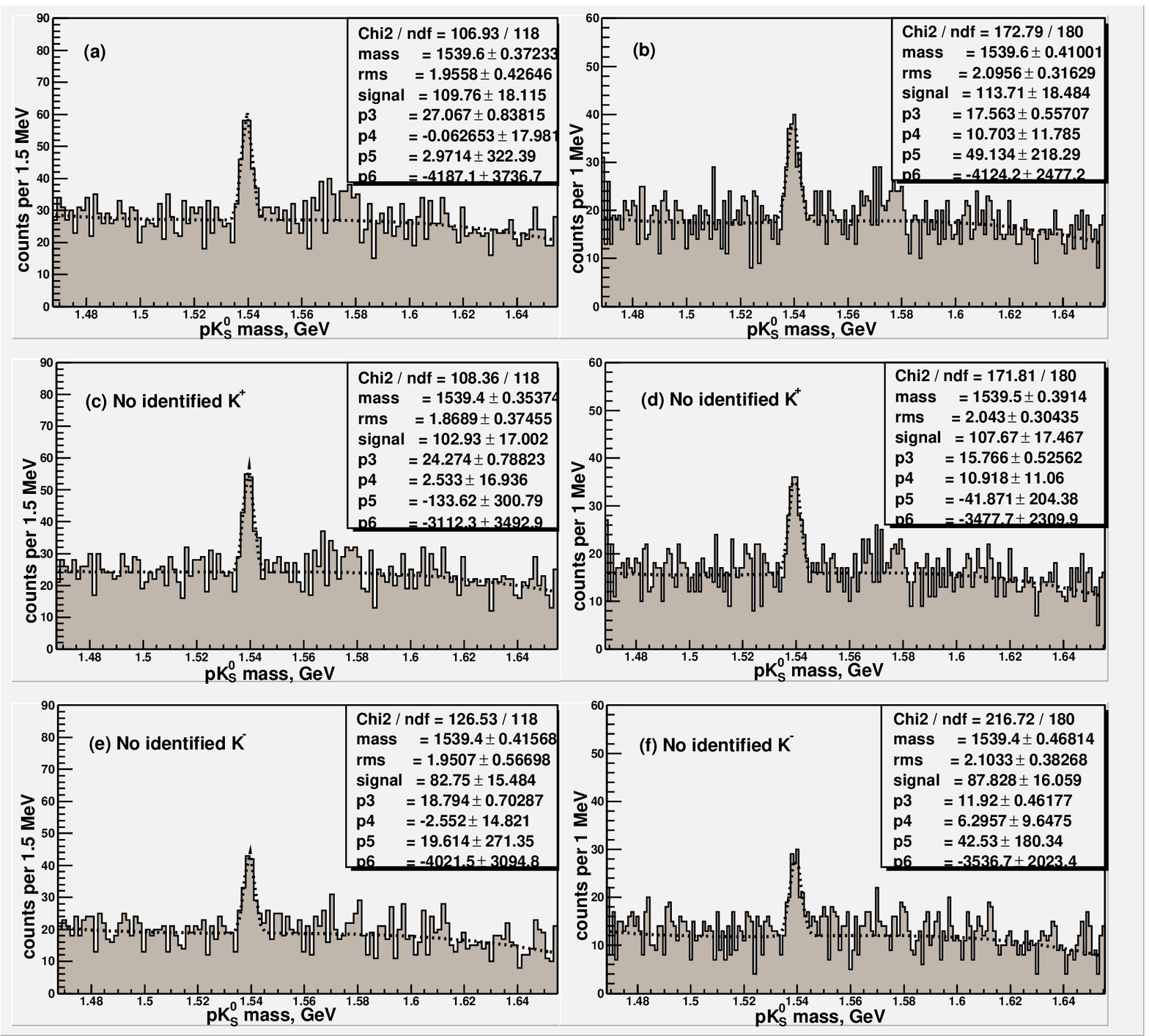}
\caption
{The $pK^0_S$ mass spectrum for $h$Cu collisions under the selection 
$x_F < 0.04$ with mass bins of 1.5 MeV (a) and 1 MeV (b). The effect
of rejecting events with identified $K^+$ mesons is shown in (c) and
(d), and of those with identified $K^-$ mesons --- in (e) and (f).}
\label{dimass-fitted}
\end{figure}

     Shown in Fig.~\ref{dimass-six-targets} is the breakdown of the
$pK^0_S$ mass spectrum for $x_F < 0.04$ to the five targets proper plus
scintillator counters (the data for the three counters have been 
combined because of small statistics). The puzzling result is that
the peak near 1540 MeV is clearly seen in the $pK^0_S$ mass spectra
for collisions in both copper targets, but not in carbon or 
scintillator. The data for collisions with copper and carbon nuclei
are compared in Fig.~\ref{dimass-vary-xF}. The low-$x_F$
enhancement near 1540 MeV is very prominent in the $pK^0_S$ mass 
spectrum for collisions in copper, whereas the spectrum for carbon 
remains regular. 
Fits of the $pK^0_S$ mass spectrum for the $h$Cu collisions under
the selection $x_F < 0.04$ are shown in Fig.~\ref{dimass-fitted}a
(bin 1.5 MeV) and Fig.~\ref{dimass-fitted}b (bin 1 MeV). The fitting
function is a Gaussian plus a third-order polynomial. The fitted
mass of the peak is near 1539 MeV, and the width is consistent with
being entirely due to instrumental resolution. (The Monte-Carlo 
prediction for a $pK^0_S$ resonance of zero intrinsic width is
$\sigma^\mathrm{MC}_m = 2.00\pm0.06$ MeV.) This allows to restrict the
intrinsic width of the observed resonance to $\Gamma < 3.0$ MeV at
90\% CL. The statistical significance of the signal, estimated as 
$S / \sqrt{B}$, is near 9$\sigma$. 

     In order to rule out a reflection from any $K^+ K^0_S$ 
resonance, we have assigned the kaon mass to proton candidates in
the $pK^0_S$ pairs with $x_F < 0.04$ selected in $h$Cu collisions.
The resultant $K^+ K^0_S$ mass distribution proves to be regular. 
In order to verify that the peak is not an artifact of detector 
acceptance for copper targets, we have (i) reflected the transverse
components of the $K^0_S$ lab momentum, $p_x \rightarrow -p_x$ and
$p_y \rightarrow -p_y$, in selected $pK^0_S$ pairs with $x_F < 0.04$,
and (ii) assigned the proton mass to identified $\pi^+$ mesons and 
then applied exactly the same selections as in Fig.~\ref{dimass-fitted}
to the $pK^0_S$ pairs formed by these ``fake" protons. In either case, 
the resultant $pK^0_S$ mass spectra prove to be featureless. 
A $pK^0_S$ peak may conceivably arise from 
the decays $\Lambda(1520) \rightarrow pK^-$ with a subsequent 
charge-exchange transition $K^- \rightarrow K^0$. However, the
simulation suggests that such a reflection from $\Lambda(1520)$ 
should occur at a lower mass (1515--1520 MeV) and have a width of 
several tens of MeV. 

     If the observed $pK^0_S$ resonance is real and indeed has positive
strangeness, a $K^+$ emitted in association with $\Theta^+$
always signals formation of an additional $s\bar{s}$ pair. Therefore, 
dropping those events that feature identified $K^+$ mesons among 
secondary particles should reduce the combinatorial background rather 
than the $\Theta^+$ peak. Likewise, a $pK^0_S$ resonance of 
negative strangeness should be emphasized by dropping events with 
identified $K^-$ mesons. We identify charged kaons as tracks with 
$p < 200$ GeV, for which the kaon probability is the highest according 
to RICH information (PID = 4). Indeed, the signal in 
Fig.~\ref{dimass-fitted} is virtually unaffected by dropping 
events with identified $K^+$ mesons, and depleted by dropping those
with identified $K^-$ mesons. This suggests that the observed $pK^0_S$
resonance has positive rather than negative strangeness. 

\begin{figure}[t]

\vspace { 7 cm }
\includegraphics{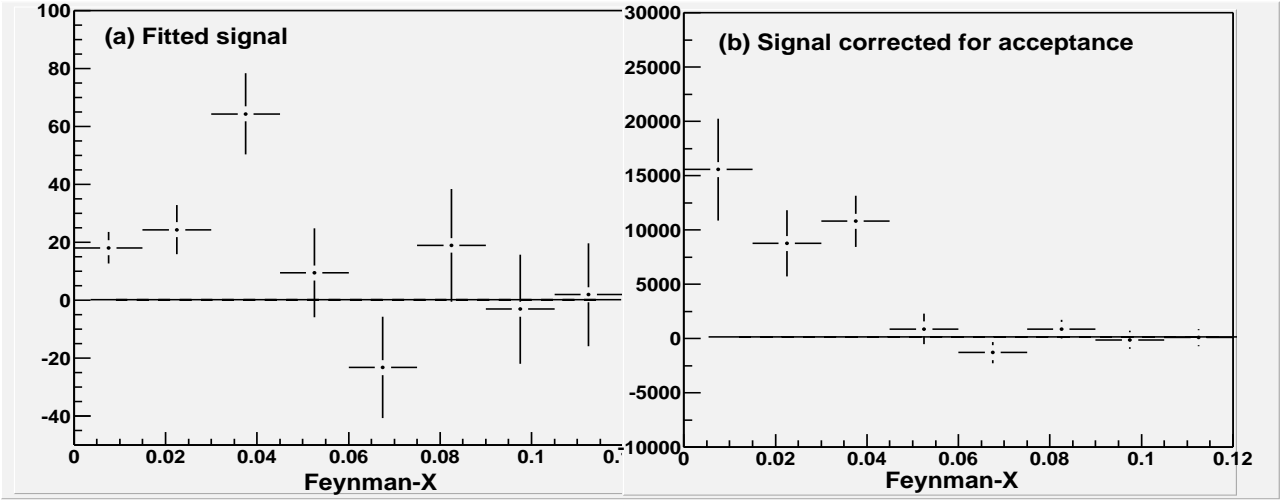}
\caption
{The fitted signal as a function of $x_F$ for collisions with copper (a).
The acceptance-corrected signal is shown in (b).}
\label{Feynman-for-copper}
\end{figure}

     In order to have a closer look at the $x_F$-dependence of the
peak, the $pK^0_S$ mass spectra for successive bins of $x_F$ are fitted
upon fixing the Gaussian's position and width to $m = 1539.5$ MeV and
$\sigma_m = 2$ MeV. The fitted signal as a function of $x_F$ is shown
in Fig.~\ref{Feynman-for-copper}a for collisions with copper.
Note that the $x_F$ dependence is strongly
distorted by detector acceptance, which steeply decreases towards 
negative values of $x_F$ where the $K^0_S$ mesons and protons are too
soft to be detected and/or identified. We estimate the acceptance as
a function of $x_F$ through an EMBED-style simulation of $\Theta^+$
production and decay, assuming isotropic angular distribution in the
$pK^0_S$ rest frame and $\langle p_\mathrm{T} \rangle = 1$ GeV. The
acceptance-corrected signal as a function of $x_F$ is shown in 
Fig.~\ref{Feynman-for-copper}b. In collisions with copper,
the acceptance-corrected number of events with 
$\Theta^+ \rightarrow pK^0_S$, $K^0_S \rightarrow \pi^+\pi^-$
is estimated as $( 3.6\pm0.6)\times10^4$ for $0 < x_F < 0.06$. 
The corresponding number for collisions with carbon is
$(0.3\pm0.7)\times10^4$. For the ratio between $\Theta^+$ yields per 
collision in carbon and copper, we obtain an upper limit
$Y^\mathrm{C}_\Theta / Y^\mathrm{Cu}_\Theta < 0.24$ at 90\% CL.

     Assuming that the branching fraction for $\Theta^+ \rightarrow pK^0$ 
is 50\%, the total number of $\Theta^+$ baryons produced with $x_F > 0$
in collisions with copper is estimated as $(2.1\pm0.4)\times10^5$.
In the same experiment, the total cross sections of $\Sigma^-$Cu and
$\pi^-$Cu collisions have been measured as $1232\pm233$ and $1032\pm179$ 
mb per nucleus, respectively \cite{Xsections}. Conservatively assuming
that all detectable $\Theta^+$ events were logged on tape and using 
$7\times10^9$ inelastic collisions with copper for reference, we obtain
$\sigma ( h\mathrm{Cu} \rightarrow \Theta^+X) = 37\pm10$ $\mu$b
for $x_F > 0$.
%  Alternatively, if the $\Theta^+$ yield per $h$Cu collision is the 
%  same for the logged and rejected events, the above estimate of the 
%  cross section should be multiplied by a factor $\sim 15$.

     To conclude, a narrow enhancement near 1539 MeV is observed 
in the  mass spectrum of the $pK^0_S$ system formed in the region of
target fragmentation ($x_F < 0.04$) in hadron collisions with copper
nuclei at 400--700 GeV. The statistical significance of the peak 
is near 9 standard deviations. Fitted width of the observed $pK^0_S$
resonance is consistent with being entirely due to experimental 
resolution, and its intrinsic width is restricted to $\Gamma < 3$ MeV
at 90\% CL in agreement with the direct measurement in 
\cite{DIANA-2014}. The data favor positive rather than negative 
strangeness for the $pK^0_S$ resonance observed in $h$Cu collisions. 
At the same time, the $pK^0_S$ mass spectrum for collisions in carbon 
is featureless. The yield of $\Theta^+$ baryons per $h$C collision
is restricted to be $< 24$\% of the yield per $h$Cu collision.

\end{document}